\documentclass[preprintnumbers, prd, twocolumn, showpacs,floatfix,preprintnumbers,
superscriptaddress, nofootinbib]{revtex4}
\usepackage{graphicx}
\usepackage{epsfig}
\usepackage{bm}
\usepackage{amssymb}
\usepackage{float}
\usepackage{amsmath}
\usepackage{subfigure}
\usepackage{dcolumn}
\usepackage{cancel}
\usepackage[colorlinks]{hyperref}
\usepackage[usenames,dvipsnames]{color}
\hypersetup{
     breaklinks=true,
    pdfstartview={FitH},    % fits the width of the page to the window
    colorlinks=true,       % false: boxed links; true: colored links
    linkcolor=blue,          % color of internal links
    citecolor=red,        % color of links to bibliography
    filecolor=magenta,      % color of file links
    urlcolor=blue,           % color of external links
    anchorcolor=green,      % Color for anchor text
    linktocpage=true
}
\newcommand{\be}{\begin{equation}}
\newcommand{\ee}{\end{equation}}
\newcommand{\bea}{\begin{eqnarray}}
\newcommand{\eea}{\end{eqnarray}}
\newcommand{\bc}{\begin{center}}
\newcommand{\ec}{\end{center}}

\begin{document}
\title{Barrow Holographic Dark Energy in non-flat Universe}
\author{Priyanka Adhikary}
\email{priyankaadhikary35@gmail.com}
\affiliation{Department of Physics, Visva-Bharati, Santiniketan -731235, India}
\author{Sudipta Das\footnote{Corresponding author}}
\email{sudipta.das@visva-bharati.ac.in}
\affiliation{Department of Physics, Visva-Bharati, Santiniketan -731235, India}
\author{Spyros Basilakos}\email{svasil@academyofathens.gr}
\affiliation{Academy of Athens, Research Center for Astronomy and Applied 
Mathematics, Soranou Efesion 4, 11527, Athens, Greece}
\affiliation{National Observatory of Athens, Lofos Nymfon, 11852 Athens, Greece}
\author{Emmanuel N. Saridakis}\email{msaridak@noa.gr}
\affiliation{National Observatory of Athens, Lofos Nymfon, 11852 Athens, 
Greece}
\affiliation{CAS Key Laboratory for Researches in Galaxies and Cosmology,
Department of Astronomy, University of Science and Technology of China, Hefei,
Anhui 230026, P.R. China}
\affiliation{School of Astronomy, School of Physical Sciences,
University of Science and Technology of China, Hefei 230026, P.R. China}

\pagestyle{myheadings}

\begin{abstract}
We construct Barrow holographic dark energy in the case of 
non-flat universe. In particular, considering closed and open spatial geometry 
we extract the differential equations that determine the evolution of the 
dark-energy density parameter, and we provide the analytical expression for the 
corresponding dark energy equation-of-state parameter. We show that  the 
scenario  can describe  the   thermal history of the universe, with  the 
sequence of matter and dark energy epochs. Comparing to the flat case,
where the phantom regime is obtained for relative large Barrow exponents, the 
incorporation of positive curvature leads the universe into the 
phantom regime for significantly smaller
values. Additionally, in the case of negative curvature
 we   find a reversed behavior, namely for increased   Barrow exponent we 
acquire algebraically higher dark-energy  equation-of-state parameters. 
Furthermore, we confront the scenario with  
   Hubble parameter measurements and supernova type Ia data.
Hence, the incorporation of slightly non-flat spatial geometry to Barrow 
holographic dark energy  improves the phenomenology 
 while keeping the new Barrow exponent to smaller 
values. 
 \end{abstract}
\maketitle

\section{Introduction}

According to the general consensus of modern cosmology, supported by a huge 
amount of cosmological observations, the Universe   experienced   accelerated 
expansion  at both early and  late times. In order to provide an explanation 
one has two main directions to follow. The first path is to introduce new forms 
of matter, such as the inflaton  \cite{Olive:1989nu,Bartolo:2004if} or the 
 dark energy   concept
\cite{Copeland:2006wr,Cai:2009zp}, while maintaining general relativity as the 
gravitational theory. The second path, is to %one can  
construct extended and modified gravitational theories, which in general give 
rise to the extra degree(s) of freedom capable of triggering  
acceleration, but still possess
 general relativity as a particular limit 
\cite{CANTATA:2021ktz,Capozziello:2011et,Cai:2015emx}. 

Nevertheless, holographic dark energy \cite{Li:2004rb,Wang:2016och} and 
holographic inflation  \cite{Nojiri:2019kkp} is an interesting  alternative 
 for the quantitative description of acceleration, that strictly speaking does 
not fall in the above two solution ways. It arises from the cosmological 
application of the   holographic 
principle 
\cite{tHooft:1993dmi,Bousso:2002ju,Fischler:1998st},
and the induced  connection between the Ultraviolet cutoff 
of a quantum field 
theory with the
largest length   
    \cite{Cohen:1998zx}, which finally  results to 
a vacuum energy of holographic origin.
Holographic dark energy   leads to interesting cosmological 
phenomenology   
\cite{Li:2004rb,Wang:2016och,Horvat:2004vn,Pavon:2005yx,
Wang:2005jx,Nojiri:2005pu,Kim:2005at,
Wang:2005ph,Setare:2008pc,Setare:2008hm}, it is in  
agreement  with   
observations
\cite{Zhang:2005hs,Li:2009bn,Feng:2007wn,Zhang:2009un,JWLee:2007JCAP,Lu:2009iv,
Micheletti:2009jy,DAgostino:2019wko,Sadri:2019qxt,Molavi:2019mlh},
and it 
has been
extended to various versions
\cite{Gong:2004fq,Saridakis:2007cy,  
Setare:2007we,Cai:2007us,Setare:2008bb,Saridakis:2007ns,Saridakis:2007wx,
Jamil:2009sq,
Gong:2009dc, 
Suwa:2009gm,BouhmadiLopez:2011xi,Chimento:2011pk,Malekjani:2012bw,
Chimento:2013se,
Khurshudyan:2014axa,
Landim:2015hqa,Pasqua:2015bfz,
Jawad:2016tne,Pourhassan:2017cba,Nojiri:2017opc,Saridakis:2017rdo,
Saridakis:2018unr,Aditya:2019bbk,Geng:2019shx,Waheed:2020cxw,Saha:2021ngs,
Saleem:2021iju}.
\par We should comment here that holographic dark energy 
models may face the causality problem \cite{Kim:2013epl}. In particular, the 
present accelerated expansion  requires the future event horizon 
to be the universe boundary    \cite{Li:2004rb}, which in turn depends on 
the future evolution of the scale factor and thus it might violate causality 
\cite{Cai:2007us}.  Nevertheless, a number of possibilities have been explored 
  to address this problem. It has been shown that suitable 
modifications of the gravitational sector in the scalar-tensor theories of 
gravity \cite{Xu:2009EPJ} or various modified  holographic models such as 
Agegraphic dark energy   \cite{Cai:2007us}, Ricci dark energy   
\cite{Gao:2009prd, Zhang:2009un} etc, can alleviate the problem through 
suitable alternative choices of the universe horizon. Additionally, there have 
been other approaches in which the causality problem can been resolved, by 
separating out the ``future-dependent'' part from the evolution equation 
\cite{Kim:2013epl}, since this  part carries the information of 
the causality violation which can be fixed by properly choosing the initial 
conditions.
\par In order to apply the holographic  principle and construct holographic dark 
energy one uses the black hole entropy expression, and thus one can obtain 
various versions of the theory  through the use of different entropies.
 Recently Barrow proposed a new black hole entropy relation that arises from 
the incorporation of  quantum-gravitational effects which may introduce  
intricate, fractal features on the black-hole area, namely 
  \cite{Barrow:2020tzx}
\begin{equation}
\label{Barrentropyy}
S_B=  \left (\frac{A}{A_0} \right )^{1+\frac{\Delta}{2}}, 
\end{equation}
with $A$   the standard horizon area ($A_0$ is the Planck area).
The new exponent   $\Delta$ lies in the range $0 \le 
\Delta \le 1$, with $\Delta=0$ corresponding to the standard smooth 
structure (in which case Barrow entropy gives back the standard 
Bekenstein-Hawking ones), and with 
$\Delta=1$ corresponding to the most intricate   structure. Hence, application 
of this extended entropy relation as the basis of holographic dark energy gives 
rise to Barrow holographic dark energy \cite{Saridakis:2020zol} which is able 
to offer improved phenomenology comparing to the standard scenarios of 
holographic dark energy 
\cite{Saridakis:2020zol,Anagnostopoulos:2020ctz,Abreu:2020cyv,Saridakis:2020lrg,
Mamon:2020spa,Abreu:2020rrh,Abreu:2020dyu,Dabrowski:2020atl,Abreu:2020wbz,
Barrow:2020kug,Srivastava:2020cyk,Das:2020rmg,Sharma:2020ylh,Pradhan:2021cbj,
Sheykhi:2021fwh,Bhardwaj:2021chg,Chakraborty:2021uzp}.\footnote{Let us 
mention   that    the form 
of the black-hole entropy   \cite{Bardeen:1973gs}   is obtained from the first 
law of black-hole thermodynamics, however this has been done under two main 
assumptions, namely that the calculations are classical and that the  theory of 
gravity is general relativity. Hence, in the literature one can find two main 
ways of extracting modified entropy relations. The first is to consider quantum 
corrections on top of classical general relativity (see e.g. 
\cite{Das:2001ic}), while the second is to consider modified theories of 
gravity, which typically lead to modified entropy relations 
\cite{Capozziello:2011et}. In 
all cases the first law of thermodynamics is valid, nevertheless it is the 
quantities that enter in it that change. } 

On the other hand, recently there is  a reheated debate on whether the spatial 
curvature of the universe is zero or not. In particular, there are arguments 
that if one considers the combined analysis of Cosmic Microwave Background 
(CMB) anisotropy power spectra of the Planck Collaboration with  the luminosity 
distance data, then  a non-flat universe is favored at
99\%  confidence level  \cite{DiValentino:2020hov}. 
Additionally   the enhanced 
lensing amplitude in the CMB power spectrum seems to 
suggest that the curvature index $k$ may 
be positive \cite{Akrami:2018vks}. 

Having these in mind, in the present work  we  are interested in constructing 
and investigating Barrow Holographic dark energy   in a non-flat universe. 
 The paper is organized as follows: In Section \ref{framework}  we     
present the basic equations for Barrow Holographic dark energy   in  both 
closed and open Friedmann-Robertson-Walker (FRW) metric. In Section 
\ref{results} we proceed to a detailed investigation of the cosmological 
behavior, focusing on  the dark energy density and equation-of-state 
parameters. In section \ref{Observational Constraints} we present the observational constraints on various parameters of the model and finally, we summarize our results  in Section 
\ref{conclusion}.

\section{Barrow holographic dark energy in non-flat geometry}
\label{framework}

In this section we desire to construct holographic dark energy in  the case of 
non-zero spatial curvature. In particular, we consider 
a non-flat  FRW line element of the form
\begin{equation}
 ds^2 = -\;dt^2 + a^2(t)\left[\frac{dr^2}{1-kr^2} +r^2 d\Omega^2 \right],
\end{equation}
where $a(t)$ is the scale factor   and $k=+1,0,-1$ corresponds to closed, flat 
and open spatial  curvature respectively.

In general, by applying Barrow entropy (\ref{Barrentropyy}) in the 
holographic framework, one obtains a holographic dark energy density of the form 
\cite{Saridakis:2020zol} 
\begin{equation}\label{rhoDE}
 \rho_{DE} = C L^{\Delta-2},
\end{equation}
with  $L$   the holographic horizon length and $C$ a 
parameter  with dimensions  $[L]^{-2-\Delta}$. Note 
that
in the case where Barrow entropy becomes the usual Bekenstein-Hawking one, 
namely for  $\Delta = 0$, expression 
(\ref{rhoDE}) gives the standard holographic dark energy $\rho_{DE} 
=C L^{-2}$  with $C=3{c^2}{M_p}^2$, where  
$c^2$ is the standard parameter of order one that is present in all holographic 
dark energy models  \cite{Li:2004rb,Wang:2016och} and $M_p$ the Planck mass.   
   
We consider that the universe is filled with the above holographic dark energy, 
as well as the matter sector.  The Friedmann 
equations  are written as
\begin{eqnarray}
3 H^2 +3\frac{k}{a^2} = {\rho}_m + {\rho}_{DE} \label{fe1}\\
2\dot{H} + 3 H^2 + \frac{k}{a^2} = -p_{DE} \label{fe2},
\end{eqnarray}
 with  $H\equiv \dot{a}/a$    the Hubble parameter, and
where $\rho_m$ is the energy density corresponding to the matter perfect fluid 
assumed to be dust,
while   $p_{DE}$ represents   the  pressure of the Barrow 
holographic dark
energy.
The two components are separately conserved, namely they obey 
\begin{eqnarray}
&&{\dot{\rho}}_m + 3 H \rho_m = 0\\
&&{\dot{\rho}}_{DE} + 3 H \left( 1+w_{DE}\right)\rho_{DE} = 0,
\label{rhoconserv}
\end{eqnarray}
where  we have introduced the dark-energy effective equation-of-state 
parameter as  
$w_{DE}\equiv\frac{p_{DE}}{\rho_{DE}}$.  Finally, it proves convenient to 
introduce
the density parameters through 
$
\Omega_{m}\equiv\frac{\rho_m}{3M_p^2H^2}$, $
\Omega_{DE}\equiv\frac{\rho_{DE}}{3M_p^2H^2}$ and $
\Omega_{k}\equiv\frac{k}{a^2H^2}$.

The last step that we need to perform is to suitably define the  largest 
length  
$L$  of the theory, namely the holographic horizon that enters in the 
definition of holographic dark energy. Although there are 
many possible choices, in the case of flat spatial geometry the most common one 
is to use 
  the   future event horizon \cite{Li:2004rb}, namely 
\begin{equation}
\label{futurehor}
R_h\equiv a\int_t^\infty \frac{dt}{a}= a\int_a^\infty \frac{da}{Ha^2}
\end{equation}
  However, if one desires to extend   holographic dark energy in a non-flat 
universe, the above length should be suitably extended 
\cite{Huang:2004ai,Setare:2006wh}. Hence, in the case of Barrow holographic 
dark energy this recipe should be followed too (note that in 
\cite{Dixit:2021phd} it was tried to apply Barrow holographic dark 
energy in a non-flat universe but with $L$ being the Hubble horizon, a choice 
that is known to be not correct \cite{Li:2004rb,Hsu:2004ri} since it cannot 
lead to acceleration). Since the corresponding extension is 
slightly different for closed and open cases, in the following subsections we 
examine them separately.

\subsection {Positive spatial curvature }

Let us start with the case of closed universe ($k=+1$).  
The horizon length $L$ is given by  $L= a r(t)$, where $r(t)$ is 
determined through \cite{Huang:2004ai,Setare:2006wh} 
\begin{equation}
\int_{0}^{r(t)} \frac{dr'}{\sqrt{1-k {r'}^2}}= 
\frac{R_h}{a }. 
\end{equation}
Thus, one obtains
\begin{equation}
\label{r}
r(t) = \frac{1}{\sqrt{k}}\sin y , 
\end{equation}
 where  
\begin{equation}
\label{ydefnit}
y=\sqrt{k} \frac{R_h}{a}  = \sqrt{k} 
\int_{x}^{\infty} \frac{dx}{a H}, 
\end{equation}
 with $x=\ln a$.
Hence, inserting  $L= a r(t)$ into (\ref{rhoDE}) we obtain
the holographic dark energy density 
\begin{equation}\label{rhoDE2}
 \rho_{DE} = C a^{\Delta-2}\left(\frac{1}{\sqrt{k}}\sin 
y\right)^{\Delta-2}.
\end{equation}

 In the following it proves convenient to use the values of the density 
parameters at present, denoted by the subscript ``0'':
\begin{eqnarray}
\label{densityparam}
\Omega_m =   \frac{\Omega_{m0}H_0^2}{a^{3}H^2},\ \ \ \ 
\Omega_k =  \frac{\Omega_{k0}H_0^2}{a^{2}H^2},
\end{eqnarray}
which in turn gives 
\begin{equation}
\frac{\Omega_k}{\Omega_m} =a \gamma ,
\end{equation}
with $\gamma\equiv \frac{\Omega_{k0}}{\Omega_{m0}} $.

Inserting (\ref{rhoDE2}) into  
(\ref{fe1}), and 
using the density parameters, we obtain 
\begin{equation}\label{aH}
\frac{1}{aH} = \frac{1}{\sqrt{{\Omega_m}_0} H_0} \ 
\left(\frac{1-\Omega_{DE}}{a^{-1} - \gamma}\right)^{\frac{1}{2}},
\end{equation}
while further insertion into (\ref{ydefnit}),(\ref{r}) leads to  
\begin{equation}
\label{Ldef11}
L=\frac{a}{\sqrt{k}} \ \sin\left[\sqrt{k} \int_{x}^{\infty} \frac{dx}{H_0 
\sqrt{\Omega_{m0}}} 
\left(\frac{1-\Omega_{DE}}{a^{-1}-\gamma}\right)^{\frac{1}{2}}\right].
\end{equation}
  On the other hand, substituting (\ref{rhoDE}) into  (\ref{fe1}), and 
using the density parameters, gives
\begin{equation}
L = \left[\frac{(1-\Omega_{DE})}{\Omega_{DE}}  \frac{C}{3M_p^2H_0^2 
\Omega_{m0}} \frac{a^{2}}{(a^{-1}-\gamma)}\right]^{\frac{1}{2-\Delta}} \;.
\label{Lsol2}
\end{equation}
Equating (\ref{Ldef11}) and (\ref{Lsol2})  one obtains the equation
\begin{equation}\label{Leqn}
    \begin{split}
\frac{a}{\sqrt{k}} \ \sin\left[\sqrt{k} \int_{x}^{\infty} \frac{dx}{H_0 
\sqrt{\Omega_{m0}}} 
\left(\frac{1-\Omega_{DE}}{a^{-1}-\gamma}\right)^{\frac{1}{2}}\right]\\  = 
\left[\frac{(1-\Omega_{DE})}{\Omega_{DE}}  \frac{C}{3M_p^2H_0^2 \Omega_{m0}}  
\frac{a^{2}}{(a^{-1}-\gamma)}\right]^{\frac{1}{2-\Delta}} .
\end{split}
\end{equation}
Differentiating equation(\ref{Leqn}) with respect to $x=\ln a$ we acquire
\begin{equation}\label{Omegapositivek}
\begin{split}
\frac{\Omega_{DE}'}{\Omega_{DE} (1-\Omega_{DE})} = \Delta + 1 + \gamma e^{x} (1-\gamma e^{x})^{-1} + \left[ Q \cos y \right.\\ \left.(\Omega_{DE})^{\frac{1}{2-\Delta}}  
(1-\Omega_{DE})^{\frac{\Delta}{2(\Delta-2)}}  e^{\frac{3 \Delta x 
}{2(\Delta-2)}}  (1-\gamma e^{x})^{\frac{\Delta}{2(2-\Delta)}}\right],
\end{split}
\end{equation}
with $$Q \equiv (2-\Delta) \left(\frac{C}{3 M_p^2}\right)^{\frac{1}{\Delta -2}} 
 \left(H_0 \sqrt{\Omega_{m0}}\right)^{\frac{\Delta}{2-\Delta}},$$   and with 
primes denoting derivatives with respect to  $x=\ln a$ .

Differential equation  (\ref{Omegapositivek}) determines the evolution of 
Barrow holographic 
dark energy for dust matter  in a closed universe. In the case where $\gamma=0$ 
(i.e. $\Omega_k=0$) it coincides with Barrow holographic 
dark energy in flat universe \cite{Saridakis:2020zol}. Additionally, in the 
case where 
  $\Delta=0$  
it coincides with the  usual holographic dark energy in a closed 
universe \cite{Huang:2004ai,Setare:2006wh}. Finally, for $\gamma=0$ and 
  $\Delta=0$ it gives back the standard holographic 
dark energy in a flat universe, namely 
 $\Omega_{DE}'|_{_{\Delta=0}}= 
\Omega_{DE}(1-\Omega_{DE})\left(1+2\sqrt{\frac{3M_p^2\Omega_{DE}}{{C}}}
\right)
$, which accepts an analytic solution (in   implicit  form) \cite{Li:2004rb}. 
 
We close this subsection by extracting the expression for the dark-energy 
equation-of-state parameter $w_{DE}$. Differentiating (\ref{rhoDE2}), using 
(\ref{ydefnit}),(\ref{r}), and inserting into (\ref{rhoconserv}), we easily 
obtain   
\begin{equation}\label{wDEpositive1}
\begin{split}
w_{DE} =- 
\left(\frac{1+\Delta}{3}\right)-\frac{Q}{3}{\left(\Omega_{DE}\right)}^{\frac{1}{
2-\Delta}} \cos y \\ \left(\frac{1-\Omega_{DE}}{1-\gamma 
e^x}\right)^{\frac{\Delta}{2(\Delta -2)}}e^{\frac{3\Delta x}{2(\Delta -2)}}.
\end{split}
\end{equation}
As expected for the flat case $\gamma=0$, equation (\ref{wDEpositive1}) 
reduces to the 
expression obtained in \cite{Saridakis:2020zol}. Moreover, for $\Delta=0$  
we acquire the expression of standard holographic dark energy in closed 
universe \cite{Huang:2004ai,Setare:2006wh}. 
Finally,  
setting  
$\gamma=0$ and 
$\Delta=0$ we re-obtain  the equation-of-state parameter for 
standard holographic dark 
energy  in flat spatial geometry \cite{Wang:2016och}.

\subsection {Negative spatial curvature }

In the case of an open universe ($k=-1$)   the horizon length $L$ is given by  
$L= a r(t)$, where $r(t)$ is 
determined through \cite{Huang:2004ai,Setare:2006wh} 
\begin{equation}
\int_{0}^{r(t)} \frac{dr'}{\sqrt{1+k {r'}^2}}= 
\frac{R_h}{a }, 
\end{equation}
leading to
\begin{equation}
\label{r22}
r(t) = \frac{1}{\sqrt{|k|}}\sinh y , 
\end{equation}
 where  
\begin{equation}
\label{ydefnit22}
y=\sqrt{|k|} \frac{R_h}{a}  = \sqrt{|k|} 
\int_{x}^{\infty} \frac{dx}{a H}, 
\end{equation}
 with $x=\ln a$.
Proceeding similarly to the previous subsection, 
we obtain 
\begin{equation}\label{sinhk}
\begin{split}
\frac{a}{\sqrt{|k|}} \ \sinh\left[\sqrt{|k|} \int_{x}^{\infty} \frac{dx}{H_0 
\sqrt{\Omega_{m0}}} 
\left(\frac{1-\Omega_{DE}}{a^{-1}-\gamma}\right)^{\frac{1}{2}}\right] \\ = 
\left[\frac{(1-\Omega_{DE})}{\Omega_{DE}} \frac{C}{3M_p^2H_0^2 \Omega_{m0}}  
\frac{a^{2}}{(a^{-1}-\gamma)}\right]^{\frac{1}{2-\Delta}} .
\end{split}
\end{equation}
Differentiating equation(\ref{sinhk}) with respect to $x=\ln a$ and using 
equation (\ref{aH}) we acquire 
\begin{equation}\label{Omeganegativek}
\begin{split}
\frac{\Omega_{DE}'}{\Omega_{DE} (1-\Omega_{DE})} = \Delta + 1 + \gamma e^{x} 
(1-\gamma e^{x})^{-1} +  \left[ Q \cosh y \right.\\ \left.(\Omega_{DE})^{\frac{1}{2-\Delta}} \ 
(1-\Omega_{DE})^{\frac{\Delta}{2(\Delta-2)}} \ e^{\frac{3 \Delta x 
}{2(\Delta-2)}} \ (1-\gamma e^{x})^{\frac{\Delta}{2(2-\Delta)}}\right]
\end{split}
\end{equation}
with $$Q \equiv (2-\Delta)\left(\frac{C}{3 M_p^2}\right)^{\frac{1}{\Delta -2}}  
\left(H_0 \sqrt{\Omega_{m0}}\right)^{\frac{\Delta}{2-\Delta}}.$$

Differential equation  (\ref{Omeganegativek}) provides the evolution of 
Barrow holographic 
dark energy for dust matter  in an open universe. In the case where $\gamma=0$ 
 it coincides with Barrow holographic 
dark energy in flat universe \cite{Saridakis:2020zol}. Furthermore, in the 
case where 
  $\Delta=0$  
it coincides with the  usual holographic dark energy in an open 
universe \cite{Huang:2004ai,Setare:2006wh}. Lastly, for $\gamma=0$ and 
  $\Delta=0$ it gives back the standard holographic 
dark energy in a flat universe    \cite{Li:2004rb}. 
 
 \begin{figure}[!]
\begin{center}
\includegraphics[width=0.84\columnwidth]{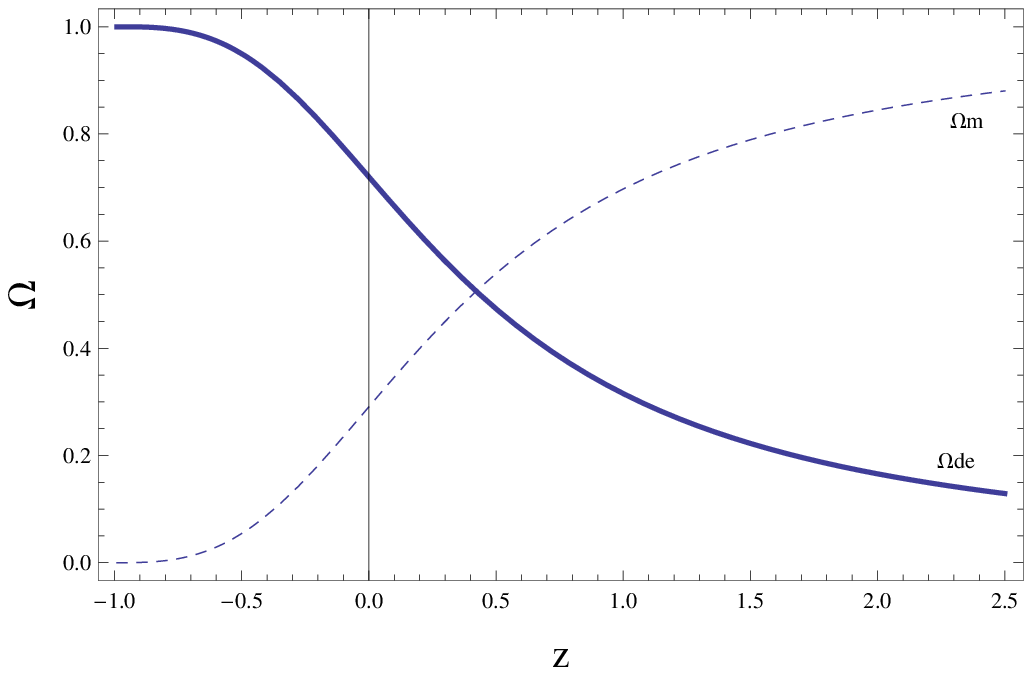}
\includegraphics[width=0.84\columnwidth]{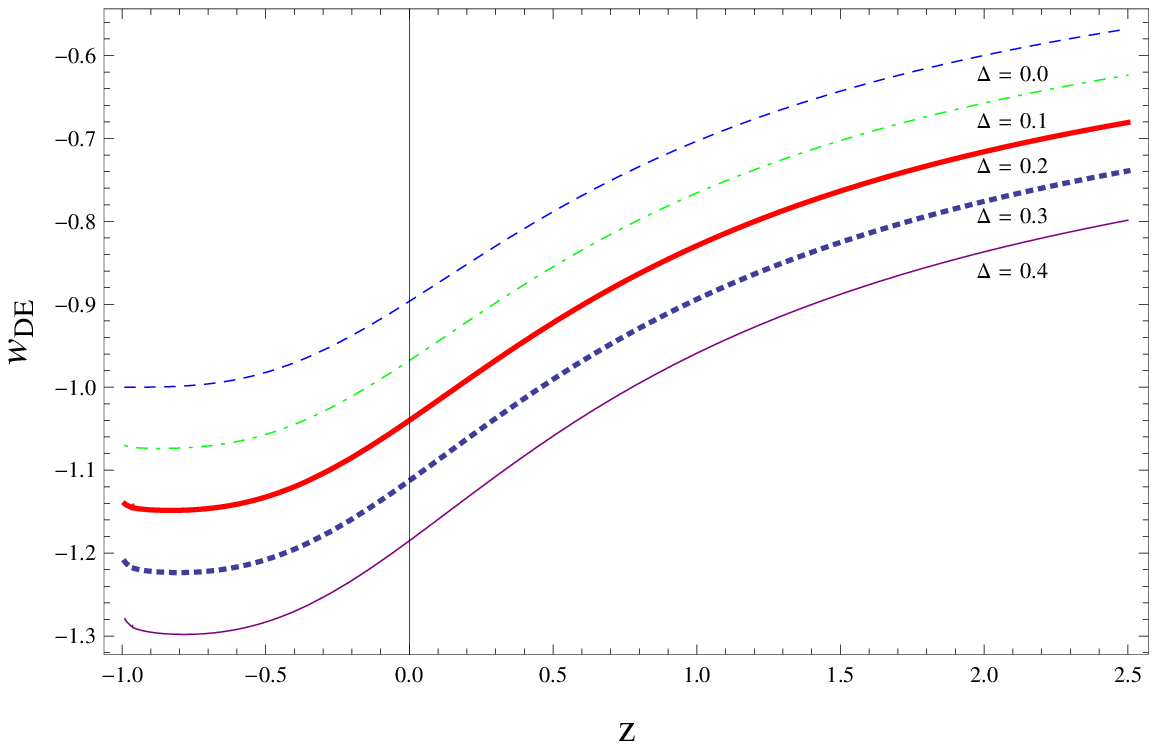}
\caption{\em 
Upper graph: The evolution of the density parameters for  matter and    Barrow 
holographic dark energy, as a function of the redshift 
$z$, in the case of a closed universe ($k=+1$), for $\Delta=0.1$ and 
${C}=3$, in    $M_p^2=1$ units. 
  Lower graph: The evolution of the   dark-energy equation-of-state 
parameter $w_{DE}(z)$, for various  $\Delta$ values.
We have imposed 
 $\Omega_{DE0}\approx0.72$, $\Omega_{m0}\approx0.27$  and $\Omega_{k0}= 
0.01$ at present. The vertical line marks the present time $z=0$. 
}
\label{figwDE}
\end{center}
\end{figure}

We proceed to extract the expression for 
the dark-energy 
equation-of-state parameter $w_{DE}$. Differentiating (\ref{rhoDE2}), using 
(\ref{ydefnit22}),(\ref{r22}), and inserting into (\ref{rhoconserv}), we easily 
obtain   
\begin{equation}\label{wDEnegative1}
\begin{split}
 w_{DE} =- 
\left(\frac{1+\Delta}{3}\right)-\frac{Q}{3}{\left(\Omega_{DE}\right)}^{\frac{1}{
2-\Delta}} \cosh y \\
\left(\frac{1-\Omega_{DE}}{1-\gamma 
e^x}\right)^{\frac{\Delta}{2(\Delta -2)}}e^{\frac{3\Delta x}{2(\Delta -2)}}.
\end{split}
\end{equation}
Similarly to the closed case, for $k=0$  equation (\ref{wDEnegative1})   
reduces to the expression for flat-universe
obtained in \cite{Saridakis:2020zol}, while with $\Delta=0$ we re-acquire the 
standard form of equation-of-state parameter for standard holographic dark 
energy     in open universe 
  \cite{Huang:2004ai,Setare:2006wh}. Lastly, for both $k=0$ and $\Delta=0$ we 
recover  
standard holographic dark 
energy  in a flat universe \cite{Wang:2016och}.

\section{Cosmological behavior} 
\label{results}

In this section  we proceed  to the investigation of the cosmological evolution 
 of Barrow holographic dark energy in closed and open universe. 
 As we mentioned above, equations (\ref{Omegapositivek}) and 
(\ref{Omeganegativek}) determine respectively the behavior of the dark-energy 
density parameter as a function of $x=\ln a$, for the two spatial-flatness 
cases. One can easily express the evolution in terms of the more 
convenient redshift, through  $x = \ln a =
-{\rm ln}(1 + z)$  (setting the current scale factor value to  $a_0 = 1$). We 
elaborate equations (\ref{Omegapositivek}) and 
(\ref{Omeganegativek})   numerically,
imposing the initial conditions 
$\Omega_m(x=-\ln(1+z)=0)\equiv\Omega_{m0} $,  
$\Omega_{DE}(x=-\ln(1+z)=0)\equiv\Omega_{DE0} $ and 
$\Omega_k(x=-\ln(1+z)=0)\equiv\Omega_{k0}$  in agreement with recent 
observations 
\cite{Akrami:2018vks}.

\begin{figure}[ht]
\begin{center}
\includegraphics[width=0.84\columnwidth]{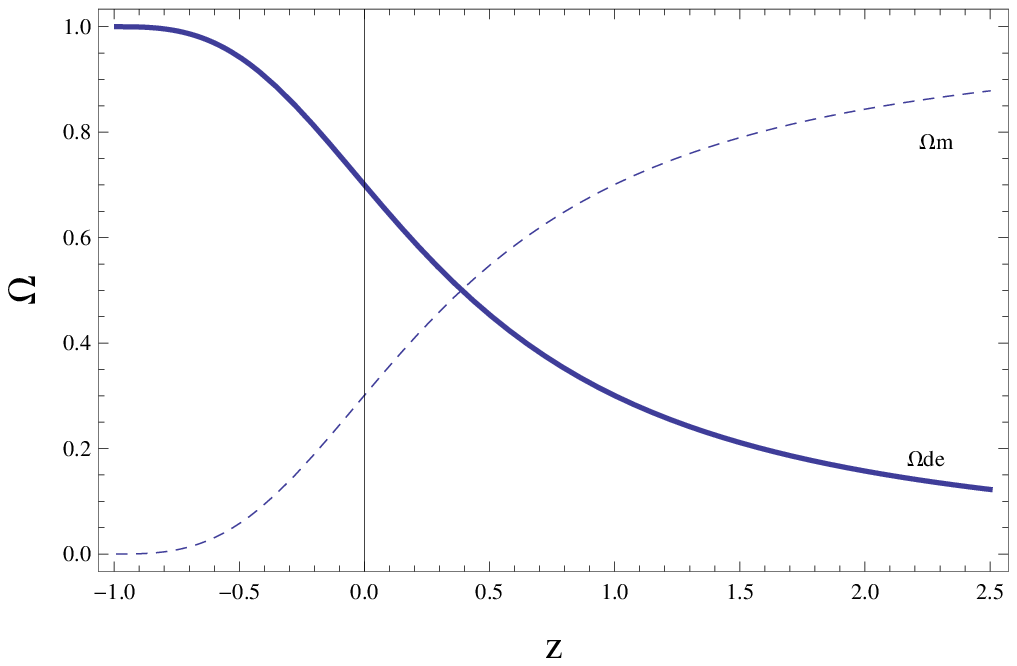}
\includegraphics[width=0.84\columnwidth]{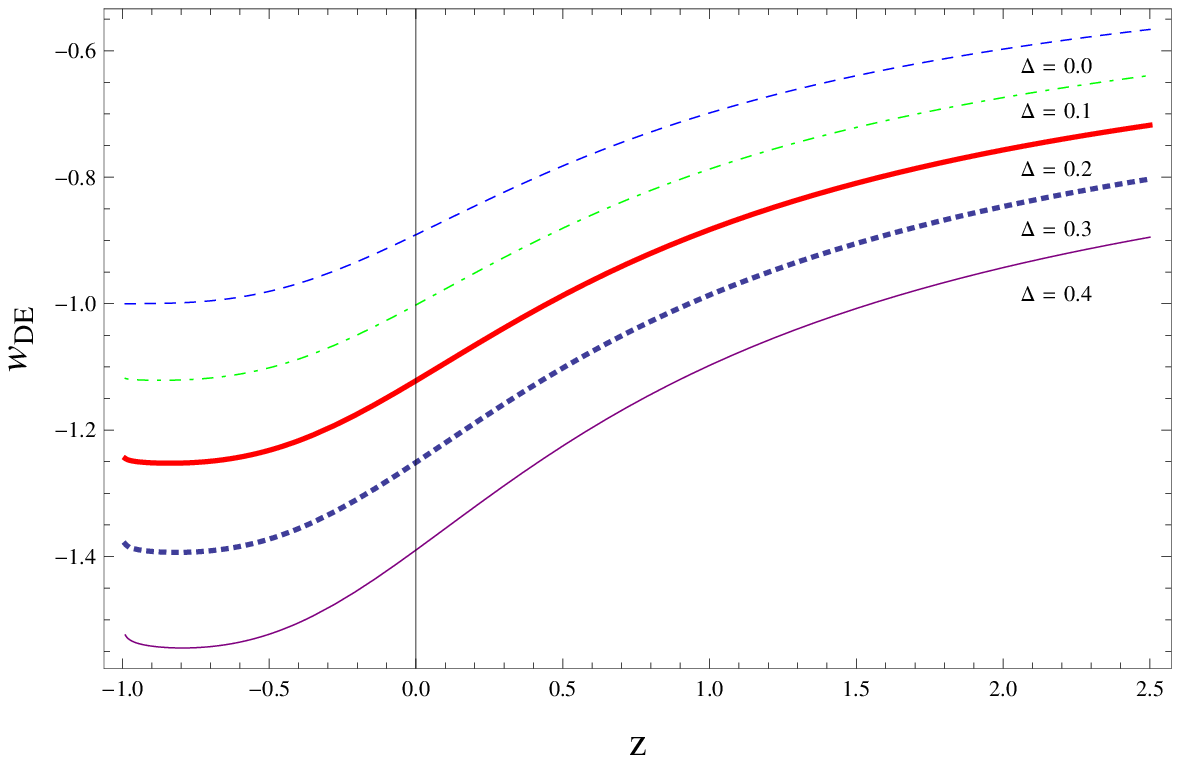}
\caption{\em
Upper graph: The evolution of the density parameters for  matter and    Barrow 
holographic dark energy, as a function of the redshift 
$z$, in the case of a closed universe ($k=+1$), for $\Delta=0.1$ and 
${C}=3$, in    $M_p^2=1$ units. 
  Lower graph: The evolution of the   dark-energy equation-of-state 
parameter $w_{DE}(z)$, for various  $\Delta$ values.
We have imposed 
 $\Omega_{DE0}\approx0.70$, $\Omega_{m0}\approx0.299$  and $\Omega_{k0}= 
0.001$ at present.  The vertical line marks the present time $z=0$. 
} 
\label{figwDElowk}
\end{center}
\end{figure}

In  the upper graph of Fig. \ref{figwDE} we depict the evolution of matter and 
dark 
energy density parameters $\Omega_{m}(z)$ and $\Omega_{DE}(z)$, in the case of 
 a  closed universe, for a given value of the Barrow exponent 
$\Delta$.   As we observe, we obtain the usual thermal history, with the 
sequence of matter and radiation epochs, with the transition from matter to dark 
energy domination happening around  $z \sim 0.4$, which is 
in agreement with the required scenario of structure formation of the universe. 
Note that for more transparency we have extended  the evolution  
up to the far future $z=-1$,  where we can see that the universe results in a 
complete dark-energy domination as expected.

\begin{figure}[ht]
\begin{center}
\includegraphics[width=0.84\columnwidth]{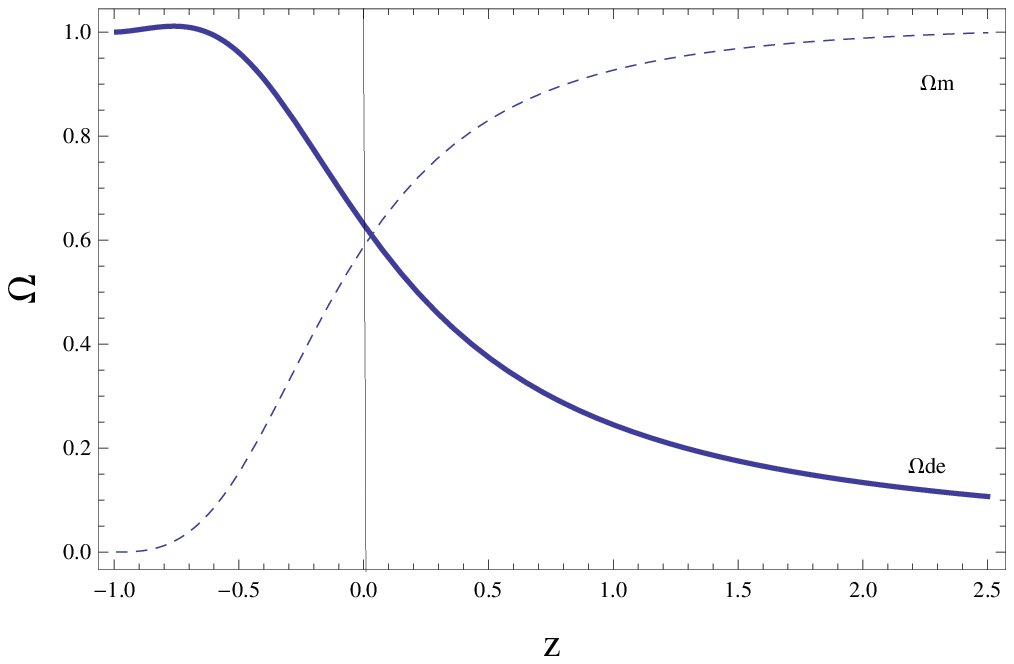}
\includegraphics[width=0.84\columnwidth]{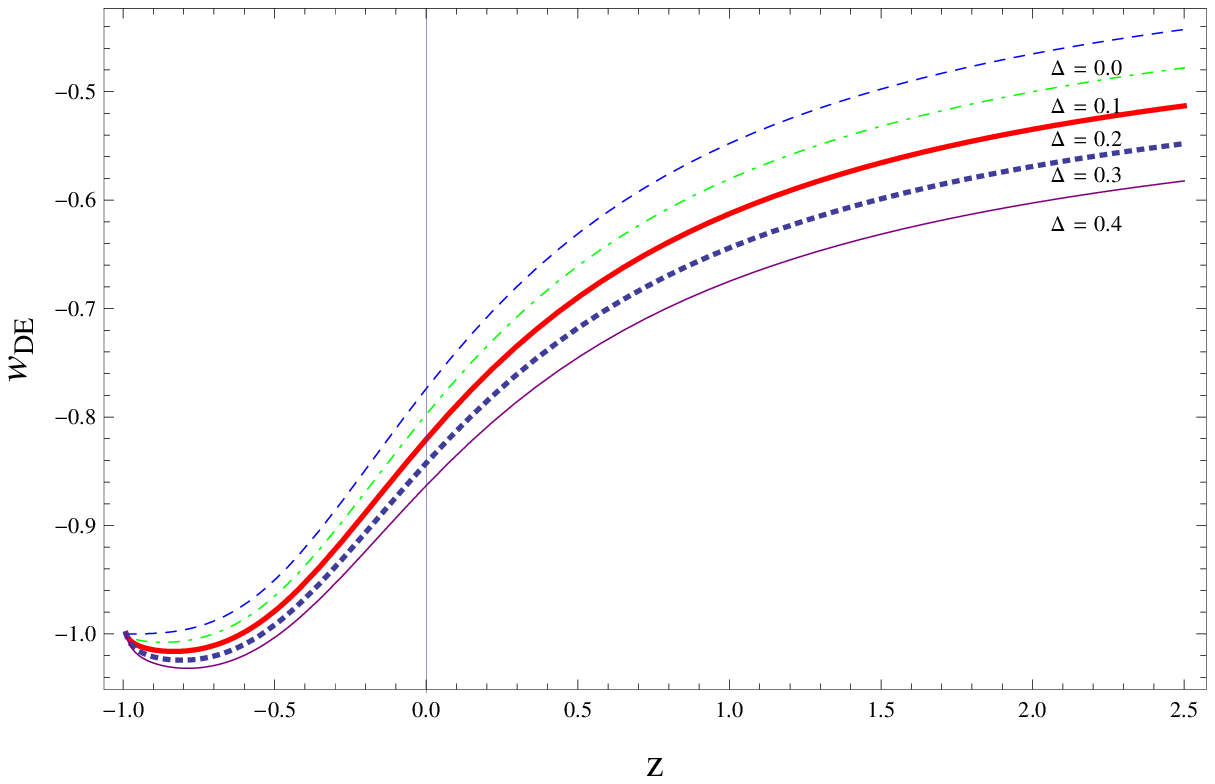}
\caption{\em 
Upper graph: The evolution of the density parameters for  matter and    Barrow 
holographic dark energy, as a function of the redshift 
$z$, in the case of a closed universe ($k=+1$), for $\Delta=0.1$ and 
${C}=3$, in    $M_p^2=1$ units. 
  Lower graph: The evolution of the   dark-energy equation-of-state 
parameter $w_{DE}(z)$, for various  $\Delta$ values.
We have imposed 
 $\Omega_{DE0}\approx0.63$, $\Omega_{m0}\approx0.27$  and $\Omega_{k0}= 
0.1$ at present. The vertical line marks the present time $z=0$.   } 
\label{figwDEhighk}
\end{center}
\end{figure}

 In order to  study in more detail the behavior of the equation-of-state 
parameter of Barrow holographic dark energy, and specifically to investigate  
how it is affected by the   exponent $\Delta$ and by $\Omega_{k0}$, in the 
lower graph of Fig. 
\ref{figwDE} we present   $w_{DE}(z)$ for the case
$k=+1$, and for various  $\Delta$ values. As we can see, 
 for   increasing 
$\Delta$ 
the   evolution of $w_{DE}(z)$ and  its current value 
$w_{DE}(z=0)\equiv w_{DE0}$  tend to obtain lower 
values. In particular, while for $\Delta=0$ the dark-energy equation-of-state 
parameter lies completely in the quintessence regime,  for $\Delta$ deviating 
from 0 the universe will result in the phantom regime, and specifically for 
$\Delta>0.03$ the phantom-divide crossing has been realized in the   past.
Hence, in the case of Barrow holographic dark energy we obtain the possibility 
to exhibit the crossing to the phantom regime, contrary to the case of 
standard holographic dark energy.

We mention here that comparing to flat Barrow holographic dark energy, in which 
the phantom regime was obtained for relative large Barrow exponents $\Delta \ge 
0.5$, the incorporation of curvature is able to drive the universe into the 
phantom regime for significantly smaller $\Delta$ values, which  is an 
advantage of the scenario since realistically one expects small Barrow 
exponents. In order to further examine the
effect of the special curvature, we 
  repeat the whole analysis for lower as well as higher values of 
$\Omega_{k0}$, and the corresponding results 
are displayed in Figs. \ref{figwDElowk}  and 
\ref{figwDEhighk}. As we observe, smaller curvature densities lead to lower 
$w_{DE}$ values, while 
the exact values of 
$\Omega_{m0}$ and $\Omega_{DE0}$ have insignificant effect. Additionally, we 
mention that in Fig. \ref{figwDEhighk}  we considered a 
non-realistically large value for $\Omega_{k0}$ in order to be able to show the 
tendency in more transparency. In particular, apart from the delay of the 
dark-energy domination (which is expected since we have imposed a lower 
$\Omega_{DE0}$), we observe that for all $\Delta$ values the universe remains 
in 
the quintessence regime, 
while in  the far future, although the phantom-divide crossing is exhibited,  
eventually all curves tend to the de Sitter phase $w_{DE}= -1$. 

\begin{figure}[ht]
\begin{center}
\includegraphics[width=0.84\columnwidth]{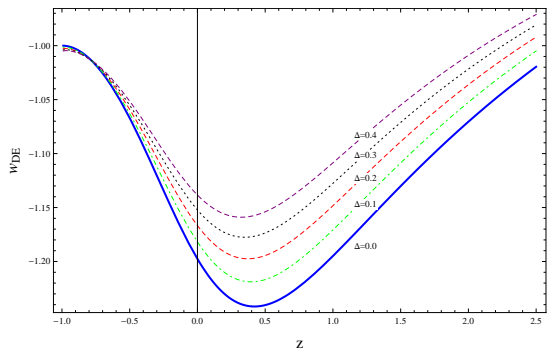}
\caption{\em
The evolution of the dark-energy equation-of-state 
parameter for   Barrow 
holographic dark energy, as a function of the redshift 
$z$, in the case of an open universe ($k=-1$), for  various  $\Delta$ 
values.
We have imposed 
 $\Omega_{DE0}\approx0.72$, $\Omega_{m0}\approx0.29$  and $\Omega_{k0}= -
0.01$ at present. The vertical line marks the present time $z=0$.  }
\label{figwDEnegativek}
\end{center}
\end{figure}

We proceed to the investigation of the negative curvature case   ($k=-1$). 
 Since the evolution of 
 $\Omega_{m}(z)$ and $\Omega_{DE}(z)$ is similar to the upper graphs of the 
previous cases, with the sequence of matter and dark-energy epochs, we omit the 
corresponding graphs and we focus on the evolution of  dark-energy 
equation-of-state 
parameter. In Fig. 
\ref{figwDEnegativek} we depict  $w_{DE}(z)$ for various  $\Delta$ values. 
Interestingly enough we now obtain a reversed behavior than in the 
positive-curvature case, namely the increased $\Delta$ leads to algebraically 
higher $w_{DE}$ values. Moreover, note that for all cases the universe is 
currently in the phantom regime, hence for the case of open spatial geometry 
 the 
phantom regime is more favorable, contrary to the case of flat universe
\cite{Saridakis:2020zol} as well as to the positive curvature case analyzed 
above. Finally, we observe the interesting behavior that  
in the far future all curves converge to the de Sitter universe, with a 
complete dark-energy domination and $w_{DE}=-1$.

 \section{Observational Constraints}
\label{Observational Constraints}

\begin{figure}[!]
\begin{center}
\includegraphics[width=0.8\columnwidth]
{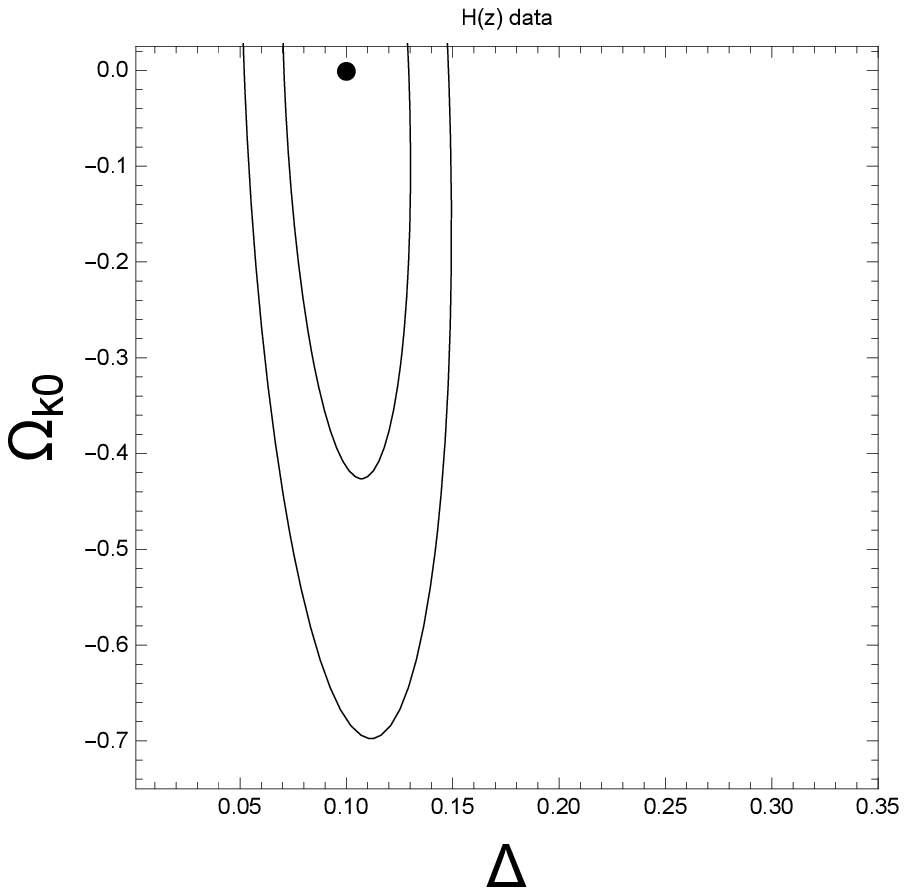} \\ 
\vspace{0.3cm}
\includegraphics[width=0.8\columnwidth]
{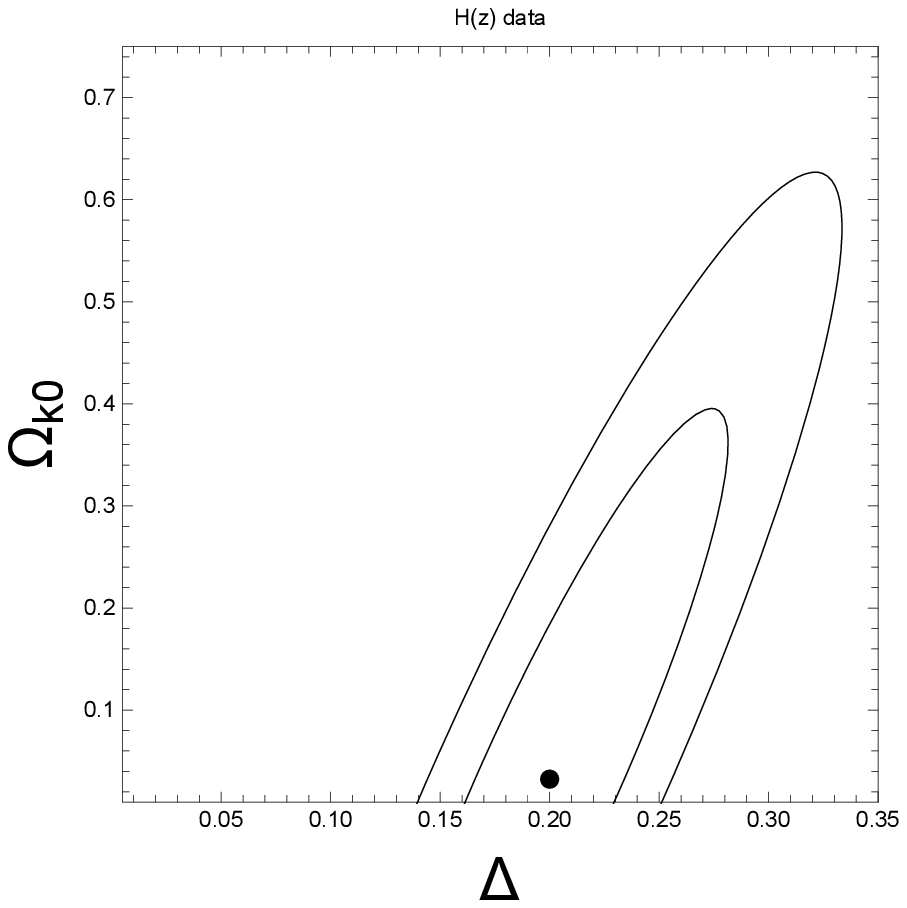}
\caption{\em 
The $1\sigma$ and $2\sigma$   iso-likelihood contours for  the scenario of  
Barrow holographic dark energy in 
non-flat universe, using Hubble data, in the case of negative curvature (upper 
panel) and positive 
curvature (lower panel).    The black dot 
represents the best fit value.}
\label{fig1}
\end{center}
\end{figure}
In this section we  proceed to the  confrontation of the 
scenario at hand with observations, 
and in particular with  Hubble measurements and  supernova type Ia  (SNIa) data.

\begin{table}[ht]
\label{bfq}
\begin{tabular}{@{}ccccc@{}}\toprule
Dataset~&$k$ &$\Delta~$& $\Omega_{k0}~$& 
 $\chi_{\text{min}}/dof$
\\  
 \colrule
%%%%%%%%%%%%%%%%%%%%%%%%%%%%%%%%%%%%%%%%%%%%%%%%%%%%%%%%%%%%%%%%%%%%%%%%%%%%%
$H(z)$ & Positive& $0.19$& $0.032$& $8.42$ \\
&Negative&$0.1$& $-0.001$& $9.24$\\
\colrule
%%%%%%%%%%%%%%%%%%%%%%%%%%%%%%%%%%%%%%%%%%%%%%%%%%%%%%%%%%%%%%%%%%%%%%%
SNIa & Positive&  $0.18$& $0.01$& $9.74$\\
&Negative&$0.21$& $-0.15$& $8.90$\\
\colrule
%%%%%%%%%%%%%%%%%%%%%%%%%%%%%%%%%%%%%%%%%%%%%%%%%%%%%%%%
$H(z)$ $+$ SNIa   & Positive&  $0.2$& $0.01$& $9.14$ \\
&Negative&$0.06$& $-0.09$& $16.23$\\
\colrule
%%%%%%%%%%%%%%%%%%%%%%%%%%%%%%%%%%%%%%%%%%%%%%%%%%%%%%%%%%%%%%%%%%%%%%%
\end{tabular} 
\caption[]{   Best-fit values of $\Delta$ and $\Omega_{k0}$ for   
 Barrow holographic dark energy in the case of 
non-flat universe, for various datasets, alongside the corresponding 
$\chi_{\text{min}}/dof$ of the 
fit, where ``dof'' stands for
degrees of freedom.} 
\label{tab:Results1}
\end{table}

For the   $H(z)$  data we  use the 29 data points of Hubble parameter 
measurements 
\cite{zhang:2014,Simon:2005,aam:2015} in the redshift range $0.07\le z\le 
2.34$. 
The corresponding $\chi^2$ function is defined as
\begin{equation}
\chi^2_{H} = \sum^{29}_{i=1}\frac{[{h}^{obs}(z_{i}) - 
{h}^{th}(z_{i})]^2}{\sigma^2_{H}(z_{i})}, \end{equation}
where ${h} = \frac{H(z)}{H_{0}}$ is the normalized Hubble parameter.   
For the SNIa dataset, we have used the Union2.1 compilation data 
\cite{Suzuki:2012} of 580 data points in the range $0.015 \le z \le 1.414$. The 
corresponding $\chi^2$   reads as \cite{NesserisPRD:2005}
\begin{equation}
\chi^2_{SN}= A - \frac{B^2}{C},
\end{equation}
with $A$, $B$ and $C$   defined as  
\begin{eqnarray}
A = \sum^{580}_{i=1} \frac{[{\mu}^{obs}(z_{i}) - 
{\mu}^{th}(z_{i})]^2}{\sigma^2_{i}},
\end{eqnarray}
\begin{eqnarray}
B = \sum^{580}_{i=1} \frac{[{\mu}^{obs}(z_i) - 
{\mu}^{th}(z_{i})]}{\sigma^2_{i}},
\end{eqnarray}
and
\begin{equation}
C = \sum^{580}_{i=1} \frac{1}{\sigma^2_{i}},
\end{equation}
where $\mu^{obs}$ represents the observed distance modulus at a particular 
redshift,  $\mu^{th}$ the corresponding theoretical value and $\sigma_{i}$ 
represents the uncertainty in the distance modulus.
Hence,  the total $\chi^{2}$ for these combined observational datasets is given 
by
\begin{equation}
\chi^{2}_{total} = \chi^{2}_{SN} +\chi^{2}_{H} .
\end{equation}

\begin{figure}[ht]
\begin{center}
\includegraphics[width=0.8\columnwidth]
{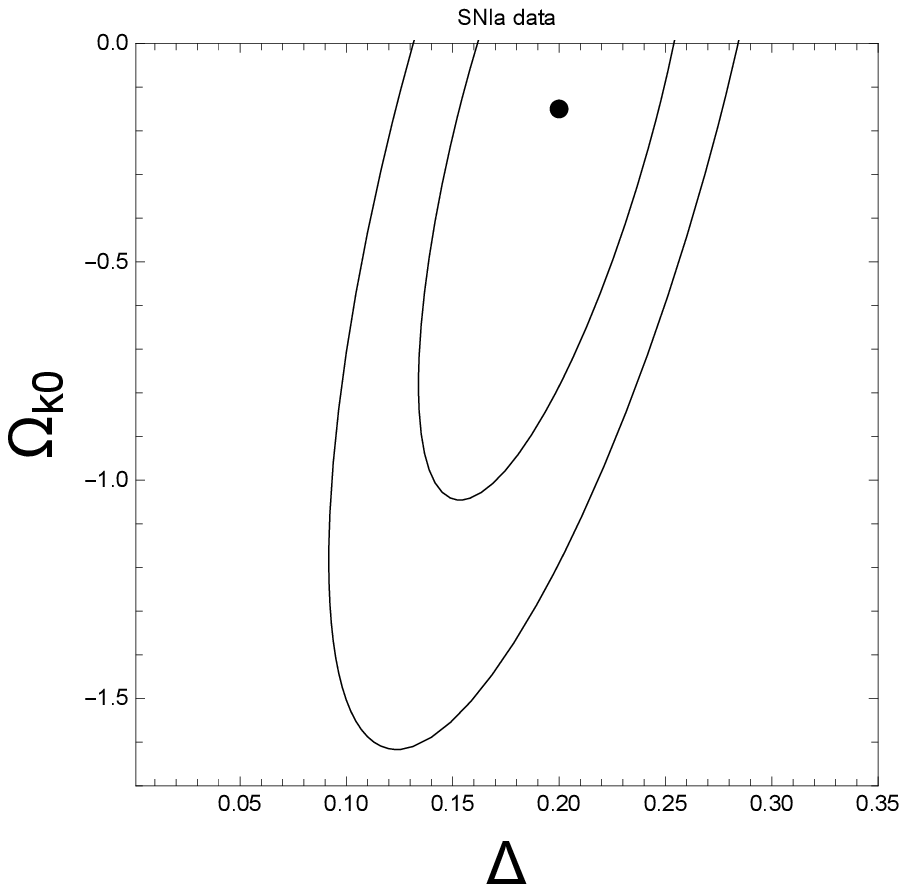} \\ \vspace{0.3cm}
\includegraphics[width=0.8\columnwidth]
{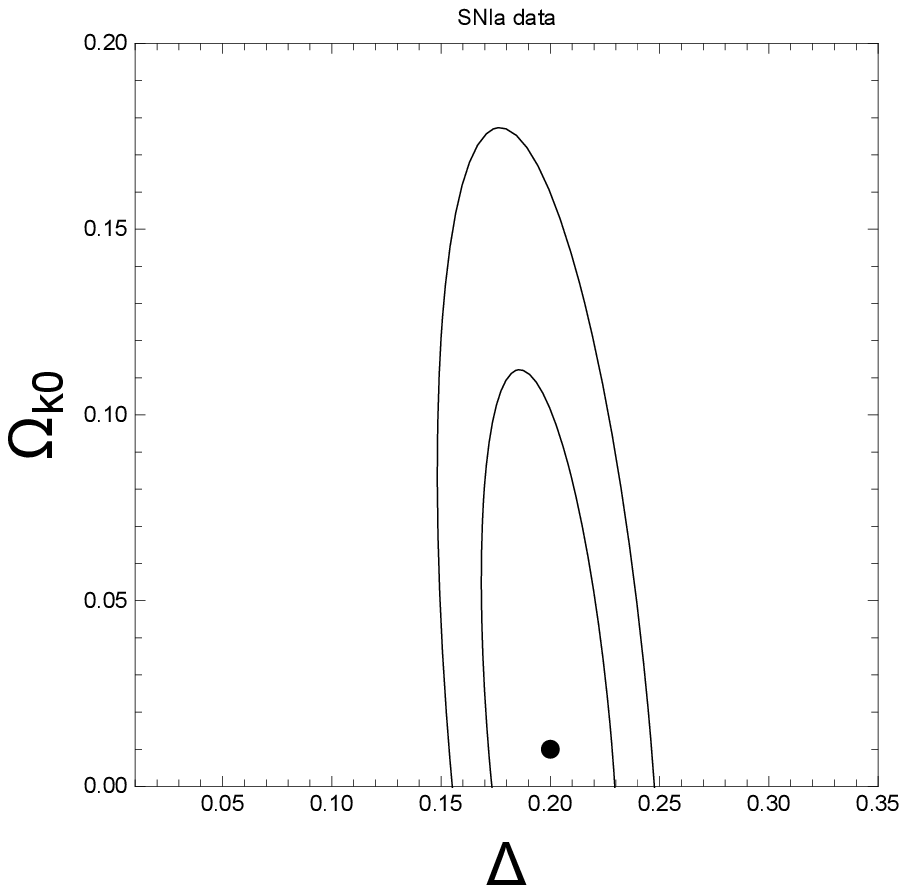}
\caption{\em  
The $1\sigma$ and $2\sigma$   iso-likelihood contours for  the scenario of  
Barrow holographic dark energy in 
non-flat universe, using SNIa data, in the case of negative curvature (upper 
panel) and positive 
curvature (lower panel).    The black dot 
represents the best fit value.}
\label{fig2}
\end{center}
\end{figure}

\begin{figure}[!]
\begin{center}
\includegraphics[width=0.8\columnwidth]
{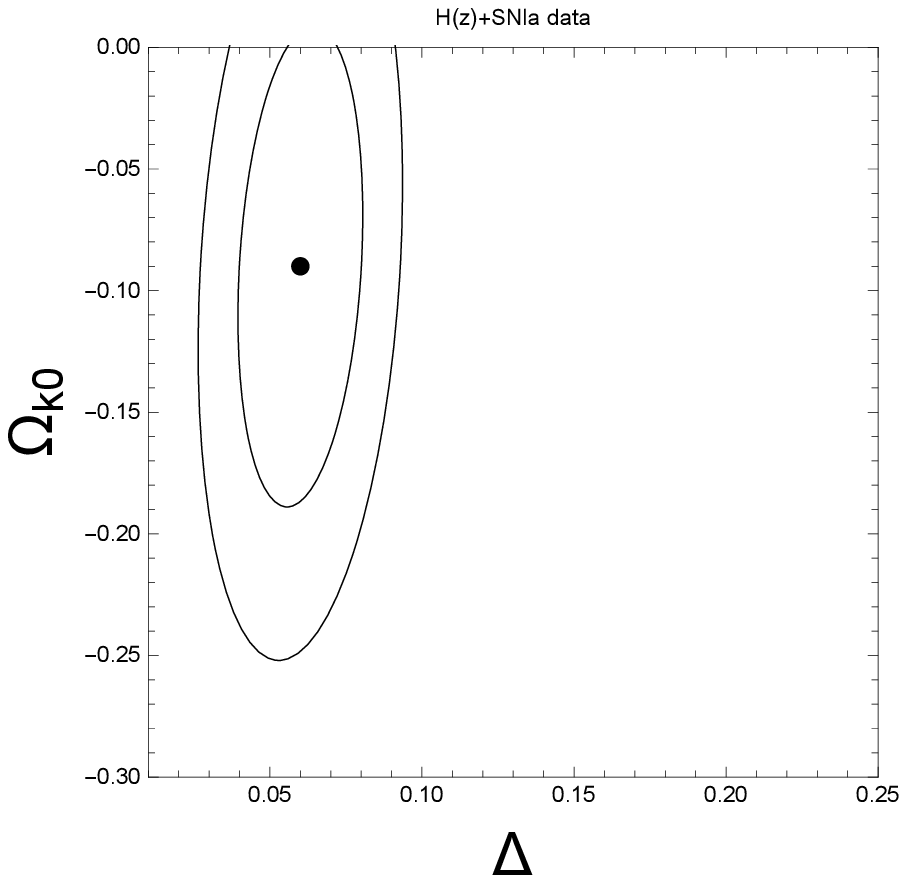} \\ \vspace{0.3cm}
\includegraphics[width=0.8\columnwidth]
{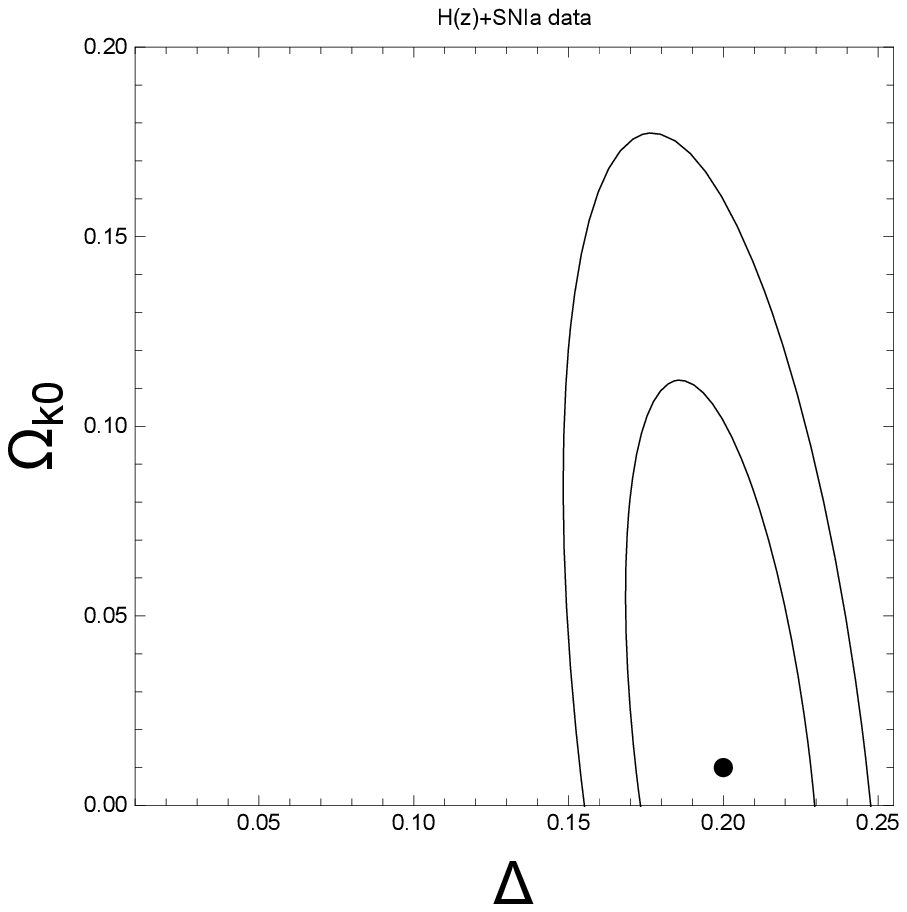}
\caption{\em  
The $1\sigma$ and $2\sigma$   iso-likelihood contours for  the scenario of  
Barrow holographic dark energy in 
non-flat universe, for the combined $H(z)$+SNIa analysis, in the case of 
negative curvature (upper 
panel) and positive 
curvature (lower panel).    The black dot 
represents the best fit value.}
\label{fig3}
\end{center}
\end{figure}

In   Table \ref{tab:Results1} we display  the resulting  best-fit values for 
the separated datasets, as well as for the combined analysis.
Additionally, in Fig. \ref{fig1} we present the $1\sigma$  and $2\sigma$ 
confidence contours in the $\Delta-\Omega_{k0}$ parameter
space in the case of Hubble data, in   Fig. \ref{fig2} in the case of SNIa 
data, and   in Fig. \ref{fig3} for the combined dataset analysis.

As we observe, in the case of negative values of $k$ the Barrow exponent $\Delta$ 
is 
constrained to smaller values, closer to standard results. Concerning    
$\Omega_{k0}$, we can see that although the best-fit values are small, comparatively larger values of $\Omega_{k0}$ are allowed at $1\sigma$ or $2\sigma$ confidence level. Furthermore, for a comprehensive analysis we have also handled 
$H_0$ as a free parameter, and the corresponding value comes out to be 
$H_0=69.86$ km s$^{-1}$ Mpc$^{-1}$, which is closer to the value obtained by 
PLANCK Collaboration. Hence, the scenario at hand might offer a way to 
alleviate the $H_0$ tension \cite{DiValentino:2020zio}. Nevertheless, we 
mention that a full investigation of this issue would require to incorporate 
additionally the CMB data and perform a joint analysis (see also 
\cite{colgain:2021,Guo:2018ans}). Such 
a full observational analysis lies beyond the scope of this first work on the 
model, and it is left for a future project.

\section{Conclusions}
\label{conclusion}

In this work we constructed Barrow holographic dark energy in the case of 
non-flat universe. The former is a holographic dark energy that arises through 
the usual application of the holographic principle in a cosmological framework, 
however it incorporates the recently proposed  Barrow entropy, instead of the 
standard  Bekenstein-Hawking one. 
Considering closed and open spatial geometry we extracted  the simple  
differential equations that determine the evolution of the dark-energy density 
parameter, and we provided the analytical expression for the corresponding 
dark energy equation-of-state parameter. 

Proceeding to the detailed investigation, we showed that the scenario at hand 
 can describe  the   thermal history of the universe, with  the sequence of 
matter and dark energy epochs. Furthermore, we examined the effect of the 
 Barrow exponent $\Delta$, as well as of the curvature density parameter at 
present, on the  dark-energy equation-of-state parameter.
As we saw,  while for $\Delta=0$ the dark-energy equation-of-state 
parameter lies completely in the quintessence regime,  for 
$\Delta>0.03$ the phantom-divide crossing has been realized in the   past, 
namely Barrow  holographic dark favors the phantom regime. 

However, the interesting feature is that 
comparing to the flat case,
where the phantom regime was 
obtained for relative large Barrow exponents $\Delta \ge 
0.5$, the incorporation of positive curvature leads the universe into the 
phantom regime for significantly smaller $\Delta$ values. This  is an 
advantage   since   one expects that only small deviations from standard 
entropy  could actually  be the 
case. Additionally, in the case of negative curvature
 we   found a reversed behavior, namely for increased $\Delta$ we obtained
algebraically higher $w_{DE}$ values, however for all cases the 
universe is currently in the phantom regime. Hence, comparing to the flat and 
closed universe,  negative curvature favors  the 
phantom regime  more intensively.
Finally, we      confronted the scenario at hand with  
   Hubble parameter measurements and supernova type Ia data, and we found that 
it can fit 
observations efficiently.
 
In summary, the incorporation of slightly non-flat spatial geometry to Barrow 
holographic dark energy improves the phenomenology 
comparing to the flat case while keeping the new Barrow exponent to smaller 
values. This is an advantage of the scenario, since in a realistic case one 
expects the Barrow exponent to be closer to the standard Bekenstein-Hawking 
value.  \\

\begin{acknowledgements}
 PA and SD acknowledges the financial support from SERB, DST, Government of 
India through the project EMR/2016/007162. SD would also like to acknowledge
IUCAA, Pune for providing support through associateship programme.  
\end{acknowledgements}

\end{document}